\documentclass[journal=jacsat,manuscript=article]{achemso}
\usepackage[version=3]{mhchem} % Formula subscripts using \ce{}
\usepackage[T1]{fontenc}       % Use modern font encodings

\author{Weijia Han}
\affiliation[OS]{Institut f\"ur Chemie neuer Materialien, Universit\"at Osnabr\"uck, Barbarastr.7, 49076 Osnabr\"uck, Germany}
\author{Peilong Hou}
\affiliation[OS]{Institut f\"ur Chemie neuer Materialien, Universit\"at Osnabr\"uck, Barbarastr.7, 49076 Osnabr\"uck, Germany}
\author{Shamaila Sadaf}
\affiliation[OS]{Institut f\"ur Chemie neuer Materialien, Universit\"at Osnabr\"uck, Barbarastr.7, 49076 Osnabr\"uck, Germany}
\author{Helmut Sch\"afer}
\affiliation[OS]{Institut f\"ur Chemie neuer Materialien, Universit\"at Osnabr\"uck, Barbarastr.7, 49076 Osnabr\"uck, Germany}
\author{Lorenz Walder}
\affiliation[OS]{Institut f\"ur Chemie neuer Materialien, Universit\"at Osnabr\"uck, Barbarastr.7, 49076 Osnabr\"uck, Germany}
\author{Martin Steinhart}
\email{martin.steinhart@uos.de}
\affiliation[OS]{Institut f\"ur Chemie neuer Materialien, Universit\"at Osnabr\"uck, Barbarastr.7, 49076 Osnabr\"uck, Germany}

\title[An \textsf{achemso} demo]
{Ordered topographically patterned silicon by insect--inspired capillary submicron stamping}

\keywords{Stamping, lithography, block copolymers, silicon, metal-assisted chemical etching, inkjet printing}

\begin{document}


%\begin{tocentry}
 %\begin{figure}[H]
%\centerline{\includegraphics{Figures/TOC.jpg}}
	%\caption{}
	%\label{TOC}
%\end{figure}
%\end{tocentry}


\begin{abstract}
Insect--inspired capillary submicron stamping and subsequent surface--limited metal--assisted chemical etching (MACE) with ammonium bifluoride as HF source is employed for high-throughput production of ordered topographically patterned silicon (tpSi). Insect feet often possess hairy contact elements through which adhesive secretion is deployed. Thus, arrays of adhesive secretion drops remain as footprints on contact surfaces. Stamps for insect--inspired capillary submicron stamping having surfaces topographically patterned with contact elements mimic the functional principles of such insect feet. They contain spongy continuous nanopore networks penetrating the entire stamps. Any ink (organic, aqueous) may be supplied from the backside of the nanoporous stamps to the contact elements. We generated ordered arrays of AgNO$_3$ dots extending mm$^2$ on Si by manual stamping with cycle times of a few seconds under ambient conditions; at higher load ordered holey AgNO$_3$ films were obtained. Surface--limited MACE correspondingly yielded either macroporous tpSi or Si pillar arrays. Inkjet printing of polymer solutions onto tpSi yielded patterns of polymer blots conformally covering the tpSi. Such blot patterns could potentially represent a starting point for the development of persistent and scratch--resistant identity labels or QR codes on silicon surfaces.             
\end{abstract}

\section{INTRODUCTION}
Topographic surface patterning of silicon components allows tailoring absorption of light for silicon-based photovoltaics \cite{SI_Hong2014,SI_Liu2014}, wettability, adhesion of coatings and the bonding strength to polymeric counterpart surfaces. Established procedures to topographically pattern silicon surfaces are typically based on combinations of top-down lithography and pattern transfer by anisotropic wet etching \cite{SI_Seidel1990a,SI_Elwenspoek1993,SI_Kelly2005} or reactive ion etching \cite{SI_Nilsson2003}. Metal-assisted chemical etching (MACE) \cite{SI_Li2000} involves the local dissolution of silicon in direct contact with certain metals in the presence of aqueous HF/H$_2$O$_2$ solutions. Arrays of silicon rods  \cite{SI_Peng2006a} and pore arrays were obtained by MACE of silicon wafers on which regularly patterned metal films or metal nanoparticle arrays had been deposited by the use of sacrificial masks or by top-down lithography. Examples for patterning methods employed to this end include laser interference lithography \cite{SI_DeBoor2010}, block copolymer lithography \cite{SI_Chang2009}, colloidal lithography \cite{SI_Huang2007} and the use of nanoporous alumina masks \cite{SI_Huang2009}. However, there is still a lack of inexpensive methods for topographic high-throughput patterning of silicon surfaces beyond costly top--down lithography or the use of sacrificial masks. Here we report the topographic patterning of silicon wafer surfaces by insect--inspired capillary submicron stamping of silver nitrate dots combined with surface--limited MACE. Thus, topographically patterned silicon (tpSi) exhibiting ordered hierarchical surface topographies -- macropores with diameters of $\sim$750 nm and depths of $\sim$200 -- $\sim$300 nm forming hexagonal arrays with a lattice constant of 1.5 $\mu$m combined with a mesoporous surface fine structure -- were obtained. Topographic patterning of silicon surfaces by insect--inspired capillary submicron stamping combined with surface--limited MACE might be advantageous in cases where alterations of the bulk properties and thus the bulk functionality of the corresponding silicon components are undesired. We show that inkjet printing of polymer solutions onto tpSi yields conformal polymer coatings. The methodology reported here might be applicable to the production of persistent arbitrary polymer patterns on silicon surfaces for potential use as permanent identification labels or QR codes.

\begin{figure}[htbp]
	\centering
		\includegraphics[width=1\textwidth]{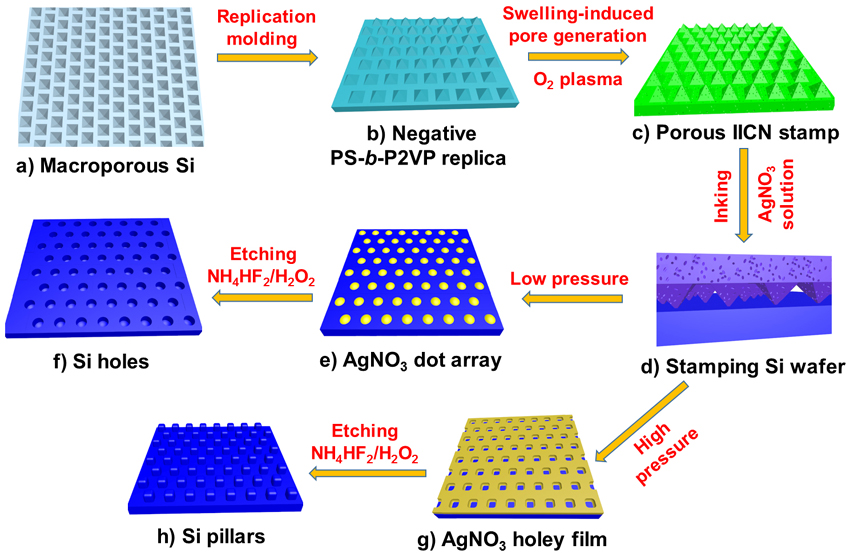}
	\caption{Topographic patterning of silicon by insect--inspired capillary submicron stamping and surface-limited MACE. a) Silanized macroporous silicon (light blue) with inverted-pyramidal pits is used as template for replication molding with PS-\textit{b}-P2VP. b) Negative PS-\textit{b}-P2VP replica of the macroporous Si templates (blue-green) are c) converted in nanoporous stamps (green) by swelling-induced pore generation and oxygen plasma treatment. d) The nanoporous stamp filled with AgNO$_3$ solution (purple) is brought into contact with a Si wafer (blue). e) If low pressure (0.04 N) is applied, AgNO$_3$ dots (yellow) are deposited at the positions of the contact elements of the nanoporous stamp. f) At the positions of the AgNO$_3$ dots macropores are generated by surface-limited MACE. g) If high pressure (0.74 N) is applied, AgNO$_3$ is deposited by inverse insect--inspired capillary submicron stamping between the contact elements of the nanoporous stamp whereas the contacts between nanoporous stamp and Si wafer are protected. Thus, a holey AgNO$_3$ film (yellow) is obtained. h) Surface-limited MACE then yields arrays of Si pillars at the positions of the contact elements of the nanoporous stamp.}
\label{IICN_scheme}
\end{figure}

\section{RESULTS AND DISCUSSION}
\subsection{Design of nanoporous stamps}
Insect adhesion is often based on the contact splitting principle \cite{IICN_Arzt2003}, which involves reversible formation of adhesive contact between insect feet and counterpart surfaces at numerous discrete contact points via a large number of hairy contact elements on the insect feet. Contact splitting is often combined with adhesion management via the deployment of liquid secretions \cite{IICN_Gorb1998}. The feet of some insects contain soft, cushion-like and spongy reservoirs (\textit{pulvilli}) storing the liquid secretion\cite{IICN_Dirks2011}. The liquid secretion may be transported through channels in the hairy contact elements to the contacts of the latter with counterpart surfaces \cite{IICN_Geiselhardt2010}. Hence, contact between hairy contact elements and counterpart surfaces is accompanied by deposition of small secretion drops at the discrete contact points.\cite{IICN_Eisner2000}. Stamps for insect--inspired capillary submicron stamping consist of a substrate mimicking the \textit{pulvilli} and of arrays of contact elements mimicking the hairy contact elements of insect feet, from which ink is supplied to counterpart surfaces. Substrate and contact elements form monolithic units which are entirely penetrated by a continuous, spongy nanopore network. In contrast to classical soft lithography with elastomeric stamps \cite{CL_Xia1998a,CL_Qin2010} and polymer pen lithography \cite{CL_Huo2008,CL_Carbonell2017} a broad range of organic or aqueous inks including dissolved functional substances can be supplied from the stamps' reverse side via the nanopore network to the contact elements or can be stored in the stamps' nanopore systems.  

\begin{figure}[htbp]
	\centering
		a) \includegraphics[width=0.6\textwidth]{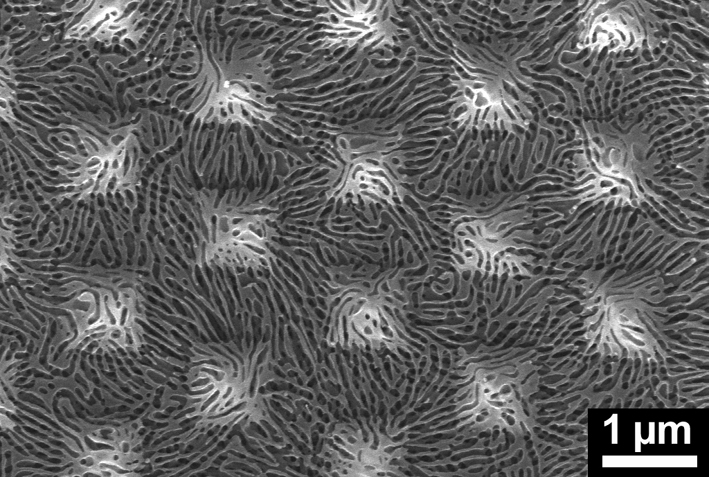}
				
		b) \includegraphics[width=0.6\textwidth]{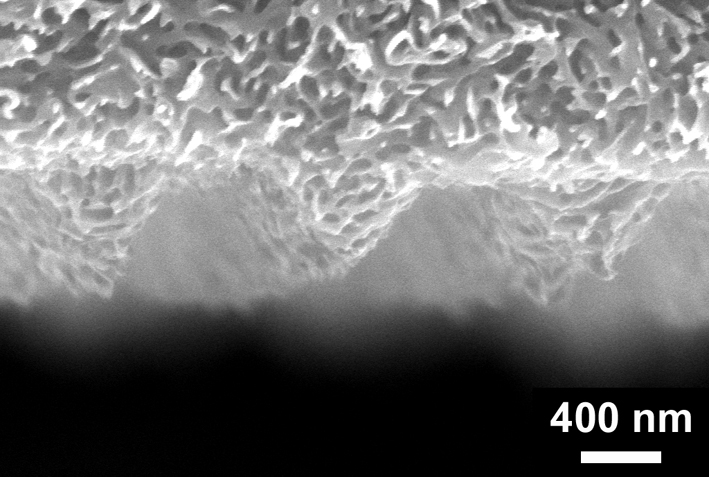}
		
	\caption{Scanning electron microscopy images of nanoporous stamps after swelling-induced pore generation and treatment with oxygen plasma. a) Top-view; b) cross-section.}
\label{stamp}
\end{figure}

Figure \ref{IICN_scheme} schematically displays the preparation of tpSi by insect-inspired capillary submicron stamping. As template for the preparation of the nanoporous stamps, we used macroporous silicon (light blue; Figure \ref{IICN_scheme}a) produced by a combination of top-down photolithography and wet etching \cite{SI_Lehmann1990,SI_Birner1998}. The macroporous silicon was patterned with hexagonal arrays of inverse-pyramidal etch pits with a nearest--neighbor distance of 1.5 $\mu$m. The base diameter of the inverse-pyramidal etch pits amounted to 920 nm and the depth to 650 nm (Supporting Information, Figure S1). First, the macroporous silicon templates were silanized. Then, the block copolymer polystyrene-\textit{block}-poly(2-vinylpyridine) PS-\textit{b}-P2VP was molded against the silanized macroporous silicon templates. After nondestructive detachment, solid PS-\textit{b}-P2VP specimens (blue-green; Figure \ref{IICN_scheme}b) with topographically patterned surfaces consisting of hexagonal arrays of pyramids (negative replicas of the inverse-pyramidal etch pits of the macroporous silicon templates) were obtained. In the next step, the topographically patterned solid PS-\textit{b}-P2VP was converted into a nanoporous stamp (green, Figure \ref{IICN_scheme}c) containing continuous sponge-like nanopore systems by swelling-induced pore generation \cite{IICN_Wang2010,IICN_Wang2016}. For this purpose, we applied a protocol established for the PS-\textit{b}-P2VP used here that yields PS-\textit{b}-P2VP specimens containing continuous nanopore systems characterized by a mean pore diameter of ~40 nm, a specific surface area of 10 m$^2$/g, and a total pore volume of 0.05 cm$^3$/g \cite{IICN_Eichler-Volf2016}. However, as obvious from Supporting Figure S2, the nanopore openings occupy only a small portion of the stamp surface. To enable efficient ink transfer, the area of the nanopore openings at the surfaces of the nanoporous stamps was increased by treatment with oxygen plasma. Thus, stamps containing nanopore systems penetrating the entire stamps (Figure \ref{stamp} and Supporting Figure S3) were obtained that consisted of a substrate with a thickness of $\sim$150 $\mu$m (Supporting Figure S3a) patterned with an array of pyramidal contact elements. The effect of the oxygen plasma treatment is obvious from a comparison of Supporting Figure S2 and Figure \ref{stamp}a.

\begin{figure}[htbp]
	\centering
		
		a) \includegraphics[width=0.4\textwidth]{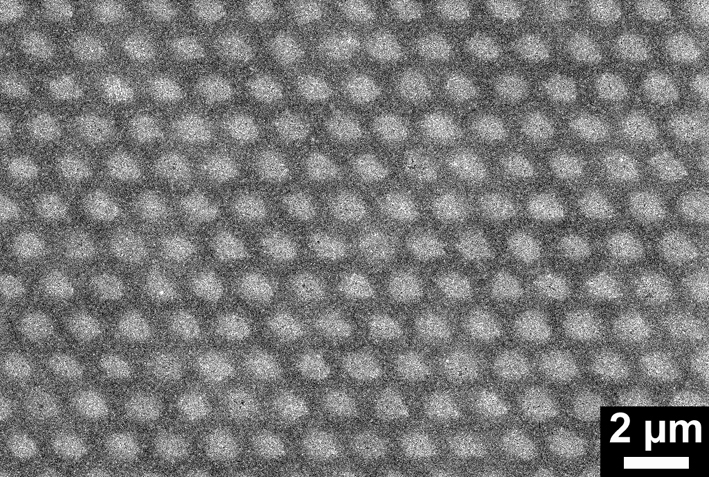}	
			
		b) \includegraphics[width=0.4\textwidth]{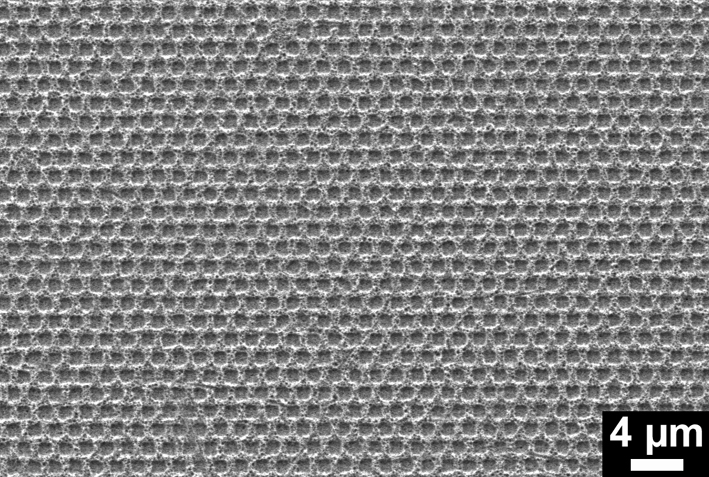}	
					
		c) \includegraphics[width=0.4\textwidth]{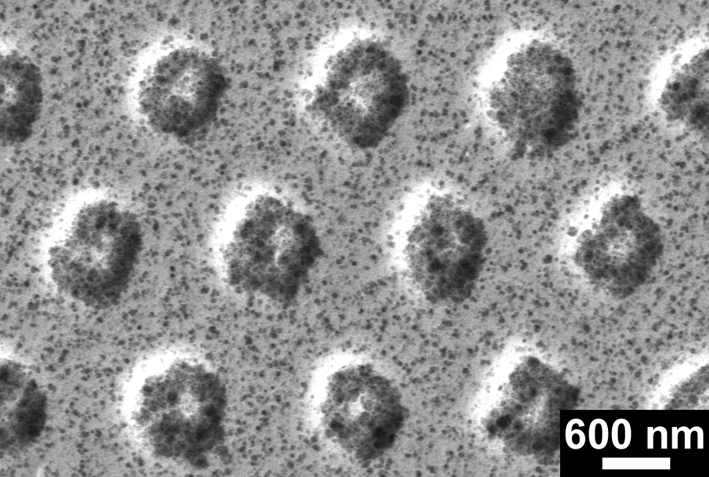}	
		
		d) \includegraphics[width=0.4\textwidth]{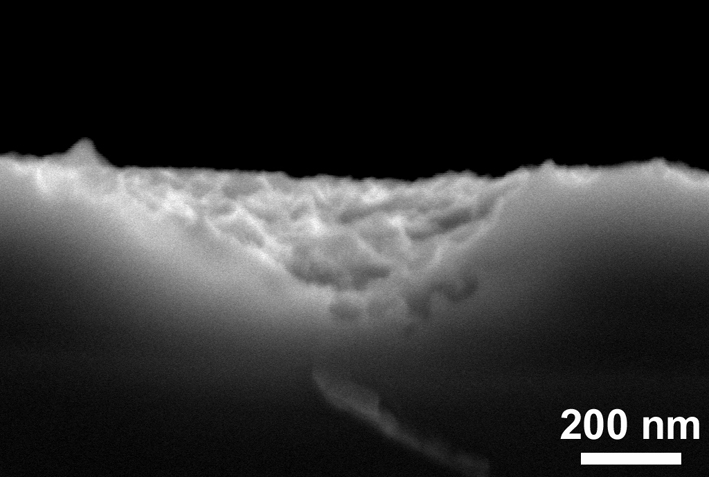}
		
\caption{Scanning electron microscopy images of a) AgNO$_3$ dots on silicon generated by insect--inspired capillary submicron stamping and b)-d) tpSi with macropores at the positions of the AgNO$_3$ dots. b),c) Top views; d) cross--sectional view.}
\label{tpSi}
\end{figure}

\subsection{Topographically patterned silicon by insect-inspired submicron stamping and MACE}
Under ambient conditions, silicon is covered by a thin native silica layer. Since MACE requires direct contact of metal catalyst and silicon, we removed the native silica layer from the Si wafers to be patterned directly prior to insect--inspired capillary submicron stamping by treatment with NH$_4$HF$_2$ solution. NH$_4$HF$_2$ is known to react with silica according the following reaction equation:\\

SiO$_2$ + 4 [NH$_4$][HF$_2$] $\rightarrow$ SiF$_4$ + 4 [NH$_4$]F + 2 H$_2$O\\

We used an aqueous solution containing AgNO$_3$ and NH$_4$HF$_2$ as ink. Nanoporous stamps saturated with ink (pink) were approached to silicon wafers (blue) treated as described above (Figure \ref{IICN_scheme}d). Applying the simple manual procedure displayed in Supporting Figure S4, the nanoporous stamp was brought into contact with the Si wafer to be stamped for $\sim$2 s under ambient conditions. As stamp holders, we used hollow poly(etheretherketone) (PEEK) cylinders that had an ink supply hole in their cylinder base. In between of cylinder base of the PEEK stamp holder and nanoporous stamp we located an elastomeric poly(dimethylsiloxane) (PDMS) interlayer with a thickness of $\sim$1.0 mm to offset misalignment between nanoporous stamp and Si wafer. Ink was supplied to the nanoporous stamp through the hole in the base of the cylindrical PEEK stamp holder and a connecting channel pierced through the elastomeric PDMS interlayer. To ensure normal approach of the nanoporous stamp to the counterpart surface, the cylindrical PEEK stamp holder fitted to the central cylindrical hole of a steel disk that was used as guiding structure (cf. Supporting Figure S4). 

As discussed below, the pressure exerted during insect-inspired capillary submicron stamping is an important parameter that determines in which way ink is deposited on the counterpart surface. The pressure was adjusted by the weight of the stamp holder (cf. Methods section). Application of a pressure of 0.04 N yielded hexagonal arrays of submicron AgNO$_3$ dots (yellow, Figure \ref{IICN_scheme}e). The AgNO$_3$ dots had diameters of $\sim$730 nm (Figure \ref{tpSi}a and Supporting Figure S5a) and heights of $\sim$ 1 nm (Supporting Figure S5b). Thus--patterend silicon wafers were converted to tpSi containing concave macropores (Figure \ref{IICN_scheme}f and Figure \ref{tpSi}b-d) at the positions of the AgNO$_3$ dots by MACE with aqueous NH$_4$HF$_2$/H$_2$O$_2$ solution. As apparent from Figure \ref{tpSi}d and Supporting Figure S6, the depth of the macropores amounted to 200-300 nm. The macropores formed hexagonal arrays with a lattice constant of 1.5 $\mu$m replicating the arrangement of the contact elements of the nanoporous stamps that extended several mm$^2$ (for large-field images cf. Supporting Figure S7 and also Supporting Figure S10a).

\begin{figure}[htbp]
	\centering
		a) \includegraphics[width=0.6\textwidth]{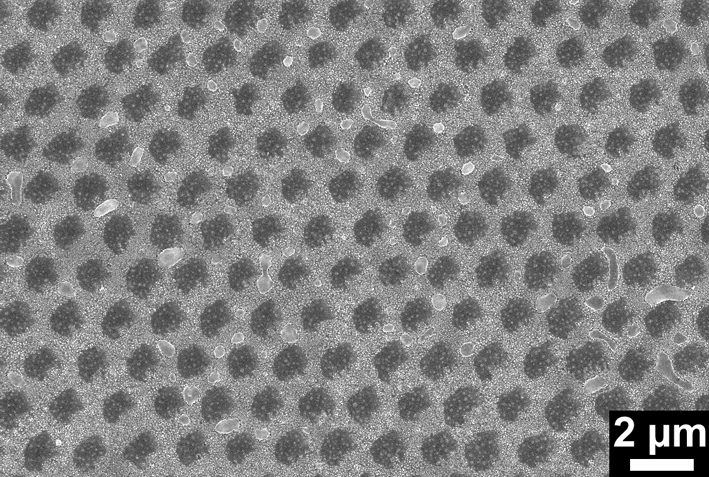}	
			
		b) \includegraphics[width=0.6\textwidth]{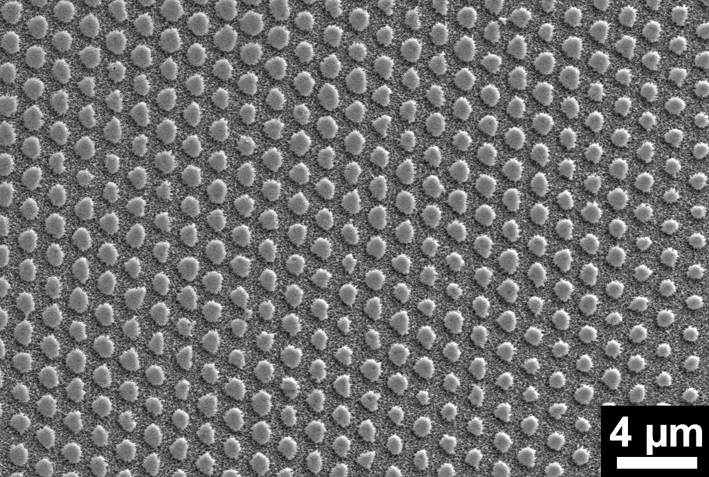}	
			
		c) \includegraphics[width=0.6\textwidth]{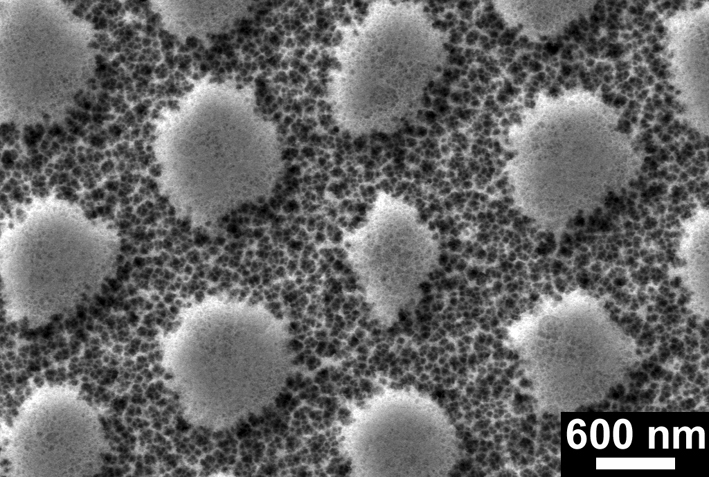}
		
	\caption{Inverse insect--inspired capillary submicron stamping at a pressure of $\sim$0.74 N. a) Scanning electron microscopy image of a Si wafer patterned with a holey AgNO$_3$ film; the AgNO$_3$ was deposited in between of the contact elements of the nanoporous stamp. b), c) Scanning electron microscopy images of tpSi with Si pillars at the positions of the contact elements of the nanoporous stamp obtained by MACE of a Si wafer patterned by inverse insect-inspired capillary submicron stamping.}
\label{tpSi_invers}
\end{figure}

Inverse insect--inspired capillary submicron stamping carried out under high pressure of 0.74 N yields holey AgNO$_3$ films on the printed Si wafers (Figures \ref{IICN_scheme}g and \ref{tpSi_invers}a; Supporting Figure S8a). Under the conditions of inverse insect--inspired capillary submicron stamping, the contact elements of the nanoporous stamps form direct solid/solid contact with the silicon wafers. Hence, no AgNO$_3$ is deposited at the positions of the contact elements. Instead, the ink is pressed out of the nanoporous stamp and covers the Si wafer around the positions of the contact elements. As a result, AgNO$_3$ films patterned with ordered arrays of holes are obtained. The holes have diameters of $\sim$600 nm and depths of $\sim$1.5 nm (Supporting Figure S8b). MACE with aqueous NH$_4$HF$_2$/H$_2$O$_2$ solution correspondingly yielded hexagonal arrays of Si pillars with diameters of $\sim$600 nm (Figure \ref{tpSi_invers}b,c and Supporting Figure S9a) and with heights of $\sim$250 nm (Supporting Figure S9b) at the positions of the holes.  

In addition to the pretreatment of the Si wafers prior to insect-inspired capillary submicron stamping the presence of NH$_4$HF$_2$ in the ink also ensures that the deposited AgNO$_3$ is in direct contact with silicon rather than with native silica on the surface of the Si wafers, as required for MACE. The MACE approach reported here comprises the replacement of aqueous HF solutions by NH$_4$HF$_2$ that can be handled easier. Moreover, no additional step to reduce the Ag(I) in the AgNO$_3$ deposited by insect--inspired capillary submicron stamping is necessary. Electroless deposition of Ag onto Si from aqueous solutions containing Ag$^+$ cations and HF is known to occur via a galvanic displacement reaction\cite{SI_Peng2006a}. While Ag$^+$ is reduced to metallic Ag, silicon is eventually oxidized to H$_2$SiF$_6$. Since the AgNO$_3$ deposited by insect--inspired capillary submicron stamping is in direct contact to Si and since both the ink and the MACE solution contain NH$_4$HF$_2$ as HF source, it is straightforward to assume that under the conditions applied here the Ag(I) is efficiently reduced. 

The surface--limited nature of the MACE process employed here can be rationalized by taking into account that i) the stability of silver catalysts under MACE conditions is limited and that ii) only small amounts of Ag are deposited by insect--inspired capillary submicron stamping of AgNO$_3$. Ag catalysts have indeed been reported to be prone to oxidation \cite{SI_Williams2017}. It has remained challenging to generate high aspect ratio features with diameters of a few 100 nm in Si by Ag-based MACE \cite{SI_Lee2008}. Hence, Ag appears to be a particularly suitable catalyst for a surface-limited MACE process without formation of MACE features characterized by high aspect ratios. A second conspicuous feature of tpSi (cf. Figures \ref{tpSi}c and \ref{tpSi_invers}c as well as Supporting Figure S10c), the ubiquitous presence of disordered mesopores, appears to be a common feature of MACE with Ag catalysts and was reported previously \cite{SI_Tsujino2005}. Excess holes that are injected into the Si contacting metallic Ag may not be immediately consumed so that they can diffuse away from the Si--Ag interface and trigger dissolution of Si elsewhere \cite{SI_Chartier2008}. Also, dissolution of Ag by the etching solution may result in non--localized etching \cite{SI_Williams2017}.

\begin{figure}[htbp]
    \centering
		
		a) \includegraphics[width=0.4\textwidth]{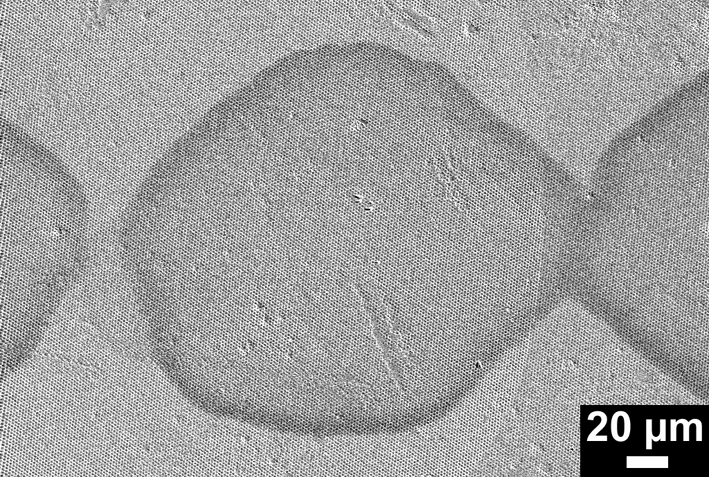}	
			
		b) \includegraphics[width=0.4\textwidth]{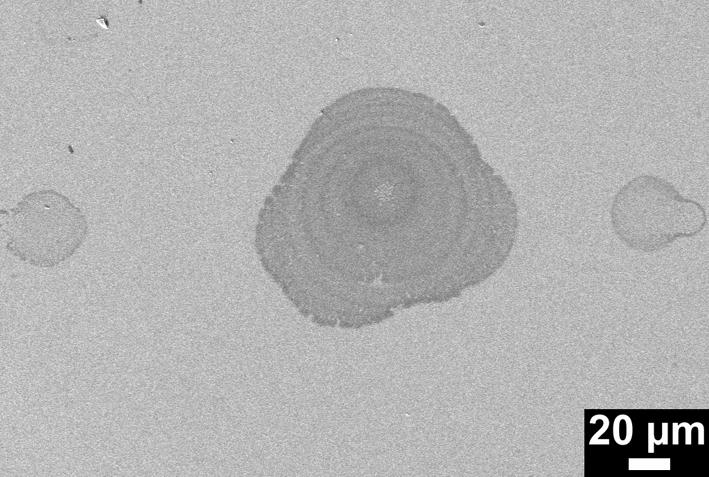}	
		
		c) \includegraphics[width=0.4\textwidth]{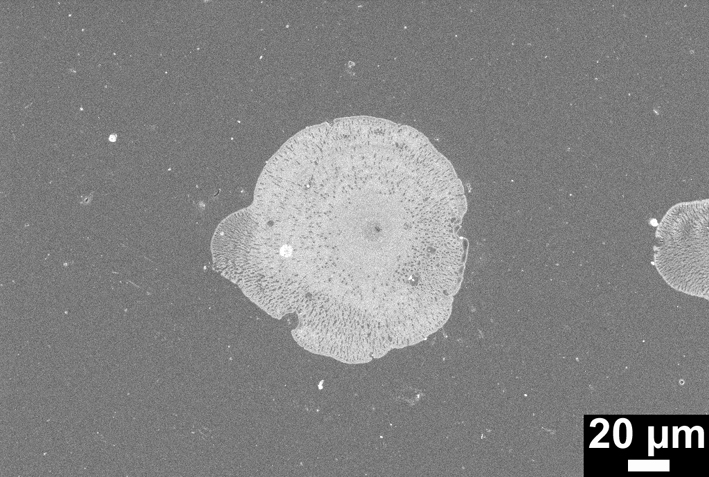}
		
  	d) \includegraphics[width=0.4\textwidth]{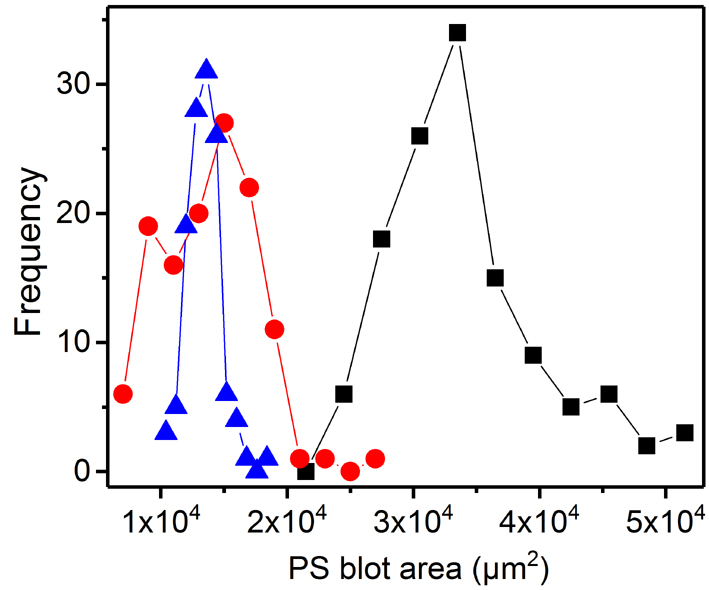}
		
		\caption{Scanning electron microscopy images of PS blots deposited by inkjet printing of 10 pL drops of PS/acetophenone solution on a) macroporous tpSi, b) mesoporous silicon and c) a smooth Si wafer. d) Frequency density of the areas of the PS blots printed on macroporous tpSi, mesoporous Si and a smooth Si wafer.}
\label{inkjet}
\end{figure}

\subsection{Inkjet printing of polymer solutions onto topographically patterned silicon}
Bonding of polymers to Si surfaces as well as the generation of persistent and scratch-resistant polymer coatings on Si surfaces have in general remained challenging. Ink jet printing \cite{BP_Tekin2008,BP_Singh2010,BP_DeGans2004a} on silicon has nevertheless been considered as particularly promising approach to combine organic electronics and silicon technology. In this context, the suppression of so--called "coffee rings"\cite{BP_Deegan1997} -- commonly unwanted ring--like deposits of non--volatile ink components -- in structures generated by inkjet printing on silicon has remained a technical problem to be solved. In exploratory inkjet printing experiments, we deposited 10 pL drops of a solution containing 1.5 wt--\% PS in acetophenone onto macroporous tpSi (Figure \ref{inkjet}a and Supporting Figure S10), onto mesoporous Si prepared in the same way than tpSi but without insect--inspired capillary submicron stamping of AgNO$_3$ (Figure \ref{inkjet}b and Supporting Figure S11) and onto smooth Si (Figure \ref{inkjet}c and Supporting Figure S12). Figure \ref{inkjet}d displays the frequencies of occurrence of blot areas obtained by evaluation of 124 PS blots per sample. The average blot area on tpSi amounted to 33929 $\mu$m$^2$, but on mesoporous Si only to 13810 $\mu$m$^2$ and on smooth Si to 13400 $\mu$m$^2$. Closer inspection of the representative PS blots on tpSi shown in Figure \ref{inkjet}a and in Supporting Figure S10 reveals that the surface topography of the tpSi is still discernible anywhere in the blot areas -- even in the outer rims (Supporting Figure S10b,c). Moreover, pronounced ring--like PS deposits were absent on tpSi. This outcome suggests that unimpeded contact line expansion occurs right after the ink drops impinge on tpSi. Apparently, vitrification of the PS due to solvent evaporation is no limiting factor for ink spreading on tpSi; it rather seems that the ink spreads until the ink film reaches a thickness close to the equilibrium value, which depends on parameters such as the disjoining pressure \cite{W_Leger1992}. A preliminary interpretation of this behavior is that contact line pinning on tpSi occurs only on a local scale confined by topographic surface features so that ink can flow around the locally pinned contact line sections. By contrast, the PS blots on mesoporous (Figure \ref{inkjet}b and Supporting Figure S11) and smooth Si (Figure \ref{inkjet}c and Supporting Figure S12) exhibit several ring--like PS deposits which might be interpreted as signatures of contact line pinning \cite{BP_Deegan1997}. If so, on mesoporous and smooth Si several pinning/depinning events occur likely already during the initial contact line expansion after impingement of the printed drops -- resulting in smaller mean areas of the PS blots than on tpSi. Moreover, the PS blots on mesoporous and smooth Si exhibit pores and voids at their surfaces that may be interpreted as signature of evaporation--induced phase separation \cite{IICN_Frank1988} resulting in the formation of PS--rich and acetophenone--rich phases. 

\section{CONCLUSIONS}
We deposited ordered metal precursor patterns by insect-inspired capillary submicron stamping on silicon wafers. In contrast to classical contact lithography, for insect--inspired capillary submicron stamping stamps entirely penetrated by spongy continuous nanopore systems are used; through the continuous spongy nanopore systems ink is supplied to the contact elements and in turn to the counterpart surfaces to be stamped. Insect--inspired capillary submicron stamping can be carried out without costly equipment under ambient conditions, allows high--throughput generation of ordered patterns using a broad range of inks and is characterized by cycle times in the second range. In the exemplary application reported here, we stamped ordered AgNO$_3$ dot arrays as well as ordered holey AgNO$_3$ films onto silicon wafers using aqueous solutions as inks. Subsequently, pattern transfer into the silicon wafers was achieved by surface--limited MACE. Silicon wafers patterned with ordered AgNO$_3$ dot arrays were thus converted into silicon topographically patterned with ordered macropore arrays extending several mm$^2$; holey AgNO$_3$ films yielded ordered Si pillar arrays. The application of the MACE catalyst Ag in the form of nm-thin AgNO$_3$ layers caused the surface-limited nature of the MACE process applied here that yields topographic surface patterns with depths of a few 100 nm rather than deep holes. Surface-limited MACE will be advantageous if it is desired to topographically pattern the surface of Si components rather than altering their bulk properties. It should also be noted that we used the solid compound NH$_4$HF$_2$ as HF source for MACE so that the handling of aqueous HF solutions could be prevented. In exploratory inkjet printing experiments we could show that polymeric inks readily spread on macroporous tpSi to form polymer blots consisting of a thin conformal coating. On mesoporous Si without insect--inspired capillary submicron stamping/MACE patterning and on smooth Si much smaller polymer blots with ring-like deposits formed that showed signatures of evaporation-induced phase separation within the ink. If permanent identity tags or QR codes need to be generated on silicon surfaces by inkjet printing or other suitable methods, persistence and scratch resistance rather than miniaturization might be the most important properties. The results reported here indicate that, in this regard, tpSi may show advantages over mesoporous Si and smooth Si.  

\section{MATERIALS AND METHODS}

\textit{Materials.} Macroporous silicon templates (Supporting Figure S1) were provided by SmartMembranes (Halle, Saale); (100)--oriented \textit{p}-type silicon wafers with a resistivity of 1-3 $\Omega$ $\cdot$ cm were purchased from Siegert Wafer. Asymmetric PS-\textit{b}-P2VP (\textit{M}$_n$(PS) = 101000 g/mol; \textit{M}$_n$(P2VP) = 29000 g/mol; \textit{M}$_w$/\textit{M}$_n$(PS-\textit{b}-P2VP) = 1.60, volume fraction of P2VP 21\%; bulk period 51 nm) was obtained from Polymer Source Inc., Canada. 1H,1H,2H,2H-Perfluorodecyltrichlorosiliane (PFDTS, purity 97\%) was purchased from ABCR GmbH. Polystyrene standard (product number 32,775-1; \textit{M}$_w$ = 29670 g/mol; \textit{M}$_n$ = 29640; M$_w$/M$_n$ = 1.01), NH$_4$HF$_2$, AgNO$_3$, HNO$_3$ (64\%-66\%), H$_2$SO$_4$ (98\%), H$_2$O$_2$ (30\%), acetophenone, tetrahydrofuran (THF, 99.9\%), and ethanol were purchased from Sigma-Aldrich.\\

\textit{Preparation of nanoporous stamps.} Macroporous silicon templates were coated with PFDTS by vapor deposition for 2 h at 85$^\circ$C and for 3 h at 130$^\circ$C following procedures reported elsewhere\cite{SI_Fadeev2000}. Then, 50 $\mu$L of a 10 wt--\% solution of PS-\textit{b}-P2VP in THF were deposited onto pieces of silanized macroporous silicon templates with areas of 1 cm$^2$. The THF was allowed to evaporate for 12 h. The obtained PS-\textit{b}-P2VP replicas of the macroporous silicon templates were non--destructively detached and then converted to nanoporous stamps with spongy-continuous nanopore systems by treatment with ethanol at 60$^\circ$C for 4 h following a protocol reported elsewhere \cite{IICN_Eichler-Volf2016}. After drying for 12 h, the smooth but porous backside of the nanoporous stamps was exposed to oxygen plasma at 100 W for 4 min. The upper surface of the nanoporous stamps, at which the pyramidal contact elements were located, was exposed to oxygen plasma at 100 Watt for 2 min. Thus, circular nanoporous stamps with an area of $\sim$0.5 cm$^2$ were obtained.\\

\textit{Patterning by insect--inspired capillary submicron stamping.} We used homemade hollow PEEK cylinders with an inner diameter of 1.0 cm and an outer diameter of 1.2 cm as stamp holders. The stamp holders were open at their upper side and had a cylinder base with a circular hole 2.0 mm in diameter for ink supply to the nanoporous stamp. To prepare the elastomeric PDMS interlayers located between nanoporous stamp and stamp holder, base and curing agent of Sylgard 184 formulation (Dow Corning, Midland, MI, USA) were mixed at a weight ratio of 10:1. The mixture was cast onto pieces of Si wafers located in plastic containers. After curing for 3 days under ambient conditions, the obtained PDMS films with a thickness of $\sim$1.0 mm were mechanically detached, roughened with P1200 SiC sanding paper on both sides and cut. Using double--sided adhesive tape (Tesa 05338), the PDMS interlayers were glued onto the cylinder bases of the cylindrical PEEK stamp holders (Supporting Figure S4), and channels for ink supply connecting the hole in the cylinder base of the stamp holders were pierced through the elastomeric PDMS interlayers with needles. Si wafer pieces with edge lengths of 1 cm were cleaned by immersion into piranha solution (98\% H$_2$SO$_4$ and 30\% H$_2$O$_2$ at a volume ratio of 3:1) at 80$^\circ$C for 30 min and copiously rinsed with deionized water. The Si wafer pieces were then dipped into a solution of 5 wt--\% NH$_4$HF$_2$ in deionized water for 2 min at room temperature to remove silica and dried in an Ar flow. Immediately thereafter, the Si wafer pieces were stamped. For this purpose, 5 $\mu$L of an aqueous solution containing 5 mmol/L AgNO$_3$ and 5 wt\% NH$_4$HF$_2$ were pipetted into the stamp holder. The ink was allowed to infiltrate the nanoporous stamp for 2 to 3 minutes. To ensure that the nanoporous stamp approaches the Si wafer piece perpendicularly, the Si wafer piece to be stamped was placed underneath a steel disk with a diameter of 5.0 cm having a perpendicular cylindrical hole, into which the cylindrical PEEK stamp holder was fitted during approach to the silicon wafer piece (Supporting Figure S4). The cylindrical PEEK stamp holders had internal screw threads at their upper openings that allowed affixing the steel cylinders with external screw threads having appropriate weights to adjust the pressure exerted during insect-inspired capillary submicron stamping.\\ 

\textit{Pattern transfer by wet--chemical etching.} To prepare tpSi, the stamped silicon wafer pieces were dipped into 40 ml of an aqueous solution containing 4.8 mol/L NH$_4$HF$_2$ and 0.4 mol/L H$_2$O$_2$ at room temperature for 40 min under exclusion of light. Subsequently, the Si wafer pieces were treated with a mixture of 20 mL 64\%-66\% HNO$_3$ and 20 ml deionized water for 5 min to remove residual AgNO$_3$.\\

\textit{Inkjet printing.} Mesoporous Si for the inkjet printing experiments was prepared in the same way as tpSi except that insect--inspired capillary submicron stamping was omitted. Si wafer pieces with an area of 1 x 1 cm$^2$ were cleaned by immersion into piranha solution (98\% H$_2$SO$_4$ and 30\% H$_2$O$_2$ at a volume ratio of 3:1) at 80$^\circ$C for 30 min and copiously rinsed with deionized water. At room temperature, the Si wafer pieces were then dipped into a solution of 5 wt--\% NH$_4$HF$_2$ in deionized water for 2 min and into 40 ml of an aqueous solution containing 4.8 mol/L NH$_4$HF$_2$ and 0.4 mol/L H$_2$O$_2$ for 40 min under exclusion of light. Inkjet printing was performed using a printer Dimatix DMP 2800 (Dimatix-Fujifilm Inc., Santa Clara, USA) equipped with a 10 pL piezoelectric printhead cartridge (DMCLCP-11610, contains 16 parallel squared nozzles with a diameter of 21 $\mu$m), which dispenses droplets with a nominal volume of 10 pL. Only one nozzle was used to print. The printer frequency was set to 5.0 kHz, the spacing between the drops was set to 250 $\mu$m. A waveform with a maximum voltage of 34 V and a pulse width of 3.5 $\mu$s was used. The temperature of the tpSi was 28$^\circ$C, and the relative humidity within the printing chamber was 30-40\%. As ink we used a solution of 1.5 wt-\% PS in acetophenone. The viscosity of the ink amounted to 1.745 cP and was measured using an Ubbelohde viscosity meter (Type No. 501 01; inner capillary diameter 0.5-0.6 mm). The prepared ink was syringe-injected into a cleaned cartridge, followed by a wait time prior to printing for 30 min. Prior to inkjet printing, all silicon samples were dipped into 5wt\% NH$_4$HF$_2$ solution for 2 min and dried in Ar flow. After printing, the patterned structures were dried at 60$^\circ$C for 2 h under ambient conditions. The areas of the PS blots generated by inkjet printing were determined with the software ImageJ using the freehand selection tool.\\

\textit{Characterization.} SEM investigations were carried out on a Zeiss Auriga microscope operated at an accelerating voltage of 7 kV. The nanoporous stamps were sputter-coated with a thin Ir-Pd alloy layer before the measurements. AFM measurements were conducted in semicontact mode with a NT-MDT NTEGRA device using GOLDEN SILICON cantilevers (NSG01/Pt; NanoLaboratory). The V-shaped cantilevers had nominal lengths of 125 $\mu$m, force constants of 1.45-15.1 N/m, and a resonance frequency of 127.8 kHz (within the range from 87 to 230 kHz). The tip curvature radius was 35 nm. The AFM images were processed by using the software Nova Px.\\ 

\section{ASSOCIATED CONTENT}
\begin{suppinfo}
Scanning electron microscopy images of macroporous silicon templates (Figure S1), of a nanoporous stamp after swelling-induced pore generation but prior to oxygen plasma treatment (Figure S2), and of cross-sections of nanoporous stamps after swelling-induced pore generation followed by oxygen plasma treatment (Figure S3). Photograph of a hollow cylindrical poly(etheretherketone) (PEEK) stamp holder onto which an intermediate elastomeric PDMS layer and a nanoporous stamp are glued (Figure S4). AFM topography images of an AgNO$_3$ dot array obtained by insect-inspired capillary submicron stamping on a Si wafer at a pressure of $\sim$0.04 N (Figure S5) and of tpSi obtained by insect-inspired capillary submicron stamping at a pressure of $\sim$0.04 N followed by MACE (Figure S6). Large-field SEM image and photograph of macroporous tpSi (Figure S7). a) AFM topography images of a holey AgNO$_3$ film obtained by inverse insect-inspired capillary submicron stamping on a Si wafer at a pressure of $\sim$0.74 N (Figure S8) and of a Si pillar array obtained by inverse insect-inspired capillary submicron stamping followed by MACE (Figure S9). SEM images of PS blots obtained by inkjet printing 10 pL drops of a 1.5 wt-\% solution of PS in acetophenone on tpSi (Figure S10), on mesoporous Si (Figure S11) and on a smooth Si wafer (Figure S12).
\end{suppinfo}

\section{AUTHOR INFORMATION}
\subsection{Corresponding Author}
*E-mail: martin.steinhart@uos.de (M. S.).

\subsection{ORCID}
Martin Steinhart: 0000-0002-5241-8498

\subsection{Notes}
The authors declare no competing financial interest.

\begin{acknowledgement}
The authors thank the European Research Council (ERC-CoG-2014, project 646742 INCANA) and the German Research Foundation (INST 190/164-1 FUGG) for funding. Support with GPC measurements by J. Kiemstedt is gratefully acknowledged.
\end{acknowledgement}

\bibliography{merged}

\newpage

\includegraphics[width=1\textwidth]{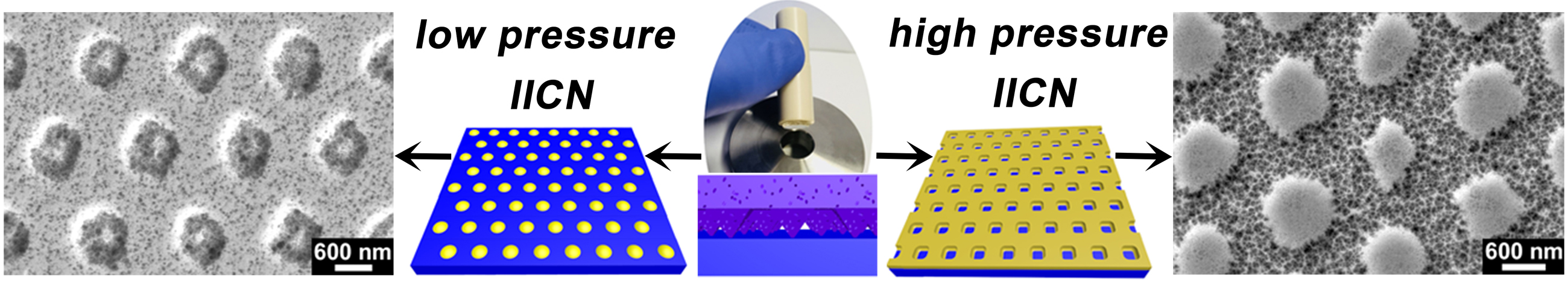}
\\
\begin{center}
Table of contents Figure
\end{center}

\end{document}